\documentclass[aip,pop]{revtex4-1}
\usepackage{graphicx}
\usepackage{multirow}
\usepackage{url}
\usepackage{color}
\begin{document}


\title{Time-resolved optical emission spectroscopic studies of picosecond laser produced Cr plasma} 



\author{Kavya H. Rao}
\affiliation{Australian Attosecond Science Facility, Centre for Quantum Dynamics, Griffith University, Nathan, Queensland-4111, Australia.}

\author{N. Smijesh}
\affiliation{Australian Attosecond Science Facility, Centre for Quantum Dynamics, Griffith University, Nathan, Queensland-4111, Australia.}
\affiliation{Department of Physics, Ume\r{a} University, Ume\r{a}, SE-901 87, Sweden}

\author{N. Klemke}
\affiliation{Australian Attosecond Science Facility, Centre for Quantum Dynamics, Griffith University, Nathan, Queensland-4111, Australia.}

\author{R. Philip}
\affiliation{Ultrafast and Nonlinear Optics Lab, Light and Matter Physics Group, Raman Research Institute, Bangalore-560080, India.}

\author{I. V. Litvinyuk}
\affiliation{Australian Attosecond Science Facility, Centre for Quantum Dynamics, Griffith University, Nathan, Queensland-4111, Australia.}

\author{R. T. Sang}
\email[]{r.sang@griffith.edu.au}
\affiliation{Australian Attosecond Science Facility, Centre for Quantum Dynamics, Griffith University, Nathan, Queensland-4111, Australia.}

\date{\today}

\begin{abstract}
Time-resolved optical emission spectroscopic measurements of a plasma generated by irradiating a Cr target using 60 picosecond (ps) and 300 ps laser pulses 
is carried out to investigate the variation in the linewidth ($\delta\lambda$) of emission from neutrals and ions for increasing ambient pressures. 
Measurements ranging from 10$^{-6}$ Torr to 10$^2$ Torr show a distinctly different variation in the $\delta\lambda$ of neutrals (Cr I) compared to that of singly ionized Cr (Cr II), for both irradiations. $\delta\lambda$ increases monotonously 
with pressure for Cr II, but an oscillation is evident at intermediate pressures for Cr I. This oscillation does not depend on the laser pulse widths used. In spite of the differences in the plasma formation mechanisms, it is experimentally found that there is an optimum intermediate background pressure for which $\delta\lambda$ of neutrals drops to a minimum. Importantly, these results underline the fact that for intermediate pressures, the usual practice of calculating the plasma number density from the $\delta\lambda$ of neutrals needs to be judiciously done, to avoid reaching inaccurate conclusions.
\end{abstract}


\maketitle 

\section{Introduction}\label{intro}
Laser produced plasmas (LPP) have gained importance in various fields owing to their diverse applications\cite{phipps2007laser} such as nanoparticle generation\cite{amoruso2005}, pulsed laser deposition\cite{Lowndes898}, high-order harmonic generation\cite{GaneevHHG}, material processing\cite{migliore1996}, wake field acceleration\cite{wakefield}, generation of EUV {\cite{EUV}} and attosecond\cite{attosecond} pulses. Detailed understanding of the plasma plume such as information about the abundance of various species and their average velocities, number density, temperature etc. are essential to devise LPP for the aforementioned applications. Even though models such as the Monte-Carlo simulation\cite{montecarlo}, blast-wave model\cite{blastwave}, drag model \cite{dragmodel}, three dimensional model \cite{threeDmodel} are available in literature for understanding plume dynamics, the transient nature of LPP makes it difficult to unravel accurate quantitative information on the plume composition and expansion. Therefore, detailed investigation of the plume using at least two of the widely accepted diagnostic techniques such as Optical Emission Spectroscopy (OES)\cite{smijesh2013OES}, Optical Time of Flight (OTOF)\cite{smijesh2014OTOF}, mass spectrometry, Langmuir probing, interferometry and shadowgraphy \cite{rao2016, griem2005principles} has to be carried out to optimise LPPs for various applications. In addition, hydrodynamics of the plasma (plume dynamics) can be understood by performing fast imaging with a time resolution of a few nanoseconds using intensified charge coupled devices (ICCDs)\cite{smijesh2014nsacceleration}. 

Among the various diagnostic techniques employed, OES is a non-invasive spectroscopic tool that uses relative intensities of characteristic line emissions from optically thin plumes for calculating the plasma temperature. Since broadening of characteristic line emissions is inevitable due to the presence of charged and uncharged entities that surround an emitter\cite{rao2016}, the emission line width can be used to estimate number density. Stark broadening, Doppler broadening, opacity broadening, broadening due to self-absorption and the effect of external magnetic fields etc. have been observed in plasmas\cite{fujimoto2004plasma, griem2005principles, Kunze}. However in LPP, Stark broadening is the major broadening mechanism\cite{Verhoff} which arises from the resultant electric field (known as the plasma microfield) generated by ions and electrons at the location of the emitter\cite{fujimoto2004plasma} (the emitter can be either an atom or an ion). 
Reports on theoretical estimations of broadening through several factors for different emitters can be found elsewhere\cite{griem2005principles, rao2016}. Most of the experiments use linewidth ($\delta\lambda$) of both neutrals and ions to calculate the number density\cite{Farid2013, Baig2006}, leading to the expected and experimentally measured result that the number density inceases with the increase in ambient pressure\cite{Farid2013, Farid2014}. However, Rao \textit{et.al.} experimentally established that for femtosecond (fs) LPP, there is an oscillation in $\delta\lambda$ of emission from neutrals for intermediate pressures when measured in the pressure range of 10$^{-6}$ Torr to 10$^2$ Torr \cite{rao2016}. In view of this observation, it was suggested that the $\delta\lambda$ of emission from ions which was found to increase monotonously with pressure (in agreement with the previous reports\cite{Farid2013, Farid2014}) would be more appropriate for calculating number density. This clearly indicates that the emission linewidth, which is an important parameter used to estimate number density\cite{smijesh2013OES}, must be carefully investigated to avoid inaccurate conclusions on the nature and dynamics of the plasma plume.

In this work, experimental investigations of chromium (Cr) plasmas generated by 60 ps and 300 ps laser pulses carried out using time-resolved OES are reported. Variation in the emission intensity ($l_{EI}$) and $\delta\lambda$ of emissions from neutrals and ions are explored for ambient pressures ranging from 10$^{-6}$ Torr to 10$^2$ Torr. Interestingly, an oscillation of the emission linewidth for neutrals is evident at intermediate pressures (0.1 Torr to 50 Torr) for both 60 ps and 300 ps laser irradiations, exhibiting a similar behaviour as observed earlier for fs laser produced Zn plasma{\cite{rao2016}}. Even though there are significant differences in the formation of fs and ps LPP, the observation reveals that there is an optimum intermediate background pressure for which the $\delta\lambda$ of neutrals drops to a minimum. Our measurements also indicate that this oscillation is independent of the pulse width of irradiation (at least for pulse durations $\leq$ 300 ps), and the material from which the plasma is generated.
\section{Experiment}\label{experiment}
Laser pulses of $\sim$ 0.6 mJ, $\sim$ 60 ps (uncompressed pulse from a multipass amplifier: \textit{Odin II, Quantronix} with a repetition rate of 1 kHz) and $\sim$ 1 mJ, $\sim$ 300 ps (uncompressed pulse from a regenerative amplifier: \textit{TSA-10, Spectra Physics} with a repetition rate of 10 Hz) at 800 nm are focussed to a spot diameter of $\sim$ 90 $\mu$m on to a 99.98\% pure Cr target (\textit{ACI Alloys Inc, USA}) to generate plasma. The fluence used in these measurements are $\sim$ 18.86 J/cm$^2$ and $\sim$ 12.74 J/cm$^2$ for 60 ps and 300 ps irradiations respectively, which is well above the threshold fluence\cite{gamaly2002ablation} $\sim$ 0.43 J/cm$^2$. Time-resolved optical emission spectra are recorded using two different high-resolution spectrometer - ICCD combinations: \textit{SP 2550 and Pi: MAX 1024f} Princeton instruments with $\sim$ 0.03 nm resolution and \textit{SR750-D1} - \textit{DH334T-18U-E3} Andor with $\sim$ 0.02 nm resolution for 60 ps and 300 ps irradiations respectively. The target is translated $\sim$ 300 $\mu$m after each ablation to avoid irradiation on the pit formed by previous ablation. Details of the experimental set up can be found elsewhere{\cite{AlSmijesh2016}}.
\section{Results and Discussion}\label{results}
Time-resolved OES measurements were carried out from 300 nm to 550 nm at 10$^{-6}$ Torr for investigating the evolution of different species and determining an appropriate integration time (or gate width, $t_w$) to perform line broadening measurements. Emission lines were identified by comparing the recorded spectra with the NIST database{\cite{NIST}}. Neutral (Cr I) lines at 396.36 nm, 396.97 nm, 425.45 nm, 427.49 nm, 428.97 nm, 520.60 nm and 520.84 nm and ionic (Cr II) lines at 312.26 nm, 312.87 and 336.66 nm are used for the analysis. The measured spectra show that the plasma plume is a mix of neutrals and ions in the earlier times of expansion for both 60 ps and 300 ps irradiaitons though their co-existence is relatively longer in the latter case. While $l_{EI}$ is used for the calculation of temperature via the Boltzmann plot method {\cite{harilal2005spectroscopic}} (or via the line-ratio method {\cite{smijesh2013OES}}), the emission line width ($\delta\lambda$) is used for the calculation of number density. The $\delta\lambda$ is given by{\cite{harilal1997carbon}}:
\begin{equation}\label{linewidth full}
\delta\lambda=2\omega\left(\frac{N_e}{10^{16}}\right)+3.5 A\left(\frac{N_e}{10^{16}}\right)\times(1-\frac{3}{4}N_D^{-1/3})\omega\left(\frac{N_e}{10^{16}}\right)
\end{equation}
where first term corresponds to the broadening caused by electron impact and the second term comes from the ion correction factor. $\omega$, $N_e$, $A$ and $N_D$ are the electron impact parameter, number density, ion broadening parameter and number of particles in the Debye sphere respectively. Assuming that stark broadening predominantly arises from electron impact, perturbations from ions can be neglected and $\delta\lambda$ can be approximated by{\cite{smijesh2013OES}}:
\begin{equation}\label{linewidth approximation}
\delta\lambda=2\omega\left(\frac{N_e}{10^{16}}\right) 
\end{equation}
from which $N_e$ can be calculated. $\delta\lambda$ can be measured by fitting a Lorentzian to an emission line.
\begin{figure}[ht!]
\includegraphics[scale=0.45]{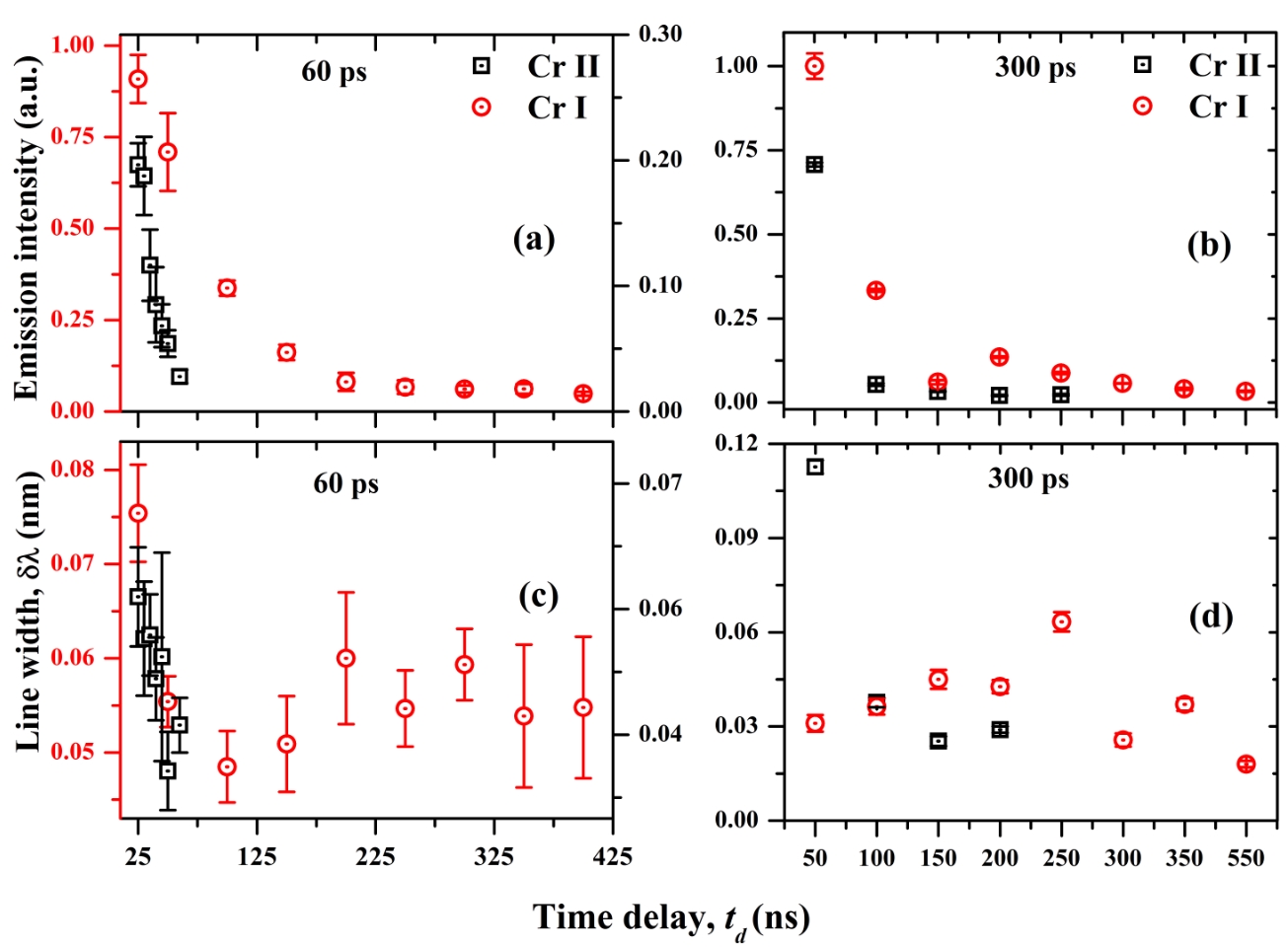}
\caption{Variation of  emission intensity and line width as function of time delay ($t_d$) for 60 ps (a,c) and 300 ps (b,d) irradiations. Ambient pressure is 10$^{-6}$ Torr. Spectral lines at 425.45 nm and 336.66 nm are measured for the Cr I and Cr II species respectively. Emission intensities are normalized to the maximum intensity for each graph (a, b). Line widths are obtained by fitting a Lorentzian function to the lines measured for each $t_d$ (c, d). Error bars are obtained from multiple measurements.}\label{GD variation-60 and 300ps}
\end{figure}
OES measurements were performed for various time delays ($t_d$s) (from 25 ns to 400 ns for 60 ps irradiation and from 50 ns to 550 ns for 300 ps irradiation.) with a fixed $t_w$ = 50 ns at a spatial position where $l_{EI}$ is maximum. The measured variation of $l_{EI}$ and $\delta\lambda$ for neutrals and ions at 10$^{-6}$ Torr as a function of $t_d$ is shown in figure \ref{GD variation-60 and 300ps}. It can be seen that emission from the plasma generated by 300 ps pulses persists for a longer time compared to that from the plasma generated by 60 ps irradiation. While emission intensity varies similarly with respect to $t_d$ (figures \ref{GD variation-60 and 300ps}a and \ref{GD variation-60 and 300ps}b) for both irradiations, $\delta\lambda$ behaves differently (figures \ref{GD variation-60 and 300ps}c and \ref{GD variation-60 and 300ps}d). The emissions from ions were recorded until $t_d$ $\leq$ 75 ns for 60 ps and $t_d$ $\leq$ 250 ns for 300 ps irradiation. Emission from neutrals is stronger than ions in both cases. These observations indicate the possibility of laser-plasma interaction as the coexistence of both neutrals and ions is longer for 300 ps irradiation. It may be recalled here that laser-plasma energy coupling occurs only if the thermalization time ($t_{th}$)\cite{IEEE-heating, Chichkov1996, gamaly2002ablation, POP-pulseduration} is shorter than the duration of the laser pulse ($\tau_p$). For Cr plasmas generated by ps laser pulses, the thermalization time and the heat diffusion time are calculated to be $\sim$ 0.35 ps and $\sim$ 40 ps \cite{gamaly2002ablation} respectively, indicating that both our laser pulses essentially act as long pulses. Clearly, the laser-plasma interaction will be longer for 300 ps excitation when compared to 60 ps excitation leading to the generation of more excited species and hence enhanced emission from ions. While a gradual decrease in the $l_{EI}$ of neutrals is evident in both cases, $l_{EI}$ of ions show a rapid decrease indicating fast recombination. $\delta\lambda_{ions}$ decrease with $t_{d}$ and the decrease is relatively sudden for 60 ps irradiation. On the other hand, $\delta\lambda_{neutrals}$ initially decrease and then show an oscillation with respect to $t_{d}$ for 60 ps irradiation, while showing negligible variation for 300 ps irradiaiton. From these observations it can be concluded that a $t_d$ $\approx$ 50 ns and $t_w$ $\leq$ 250 ns would be a suitable choice for line broadening investigations owing to the co-existence of both ions and neutrals in this region.
\begin{figure}[ht!]
\includegraphics[scale=0.45]{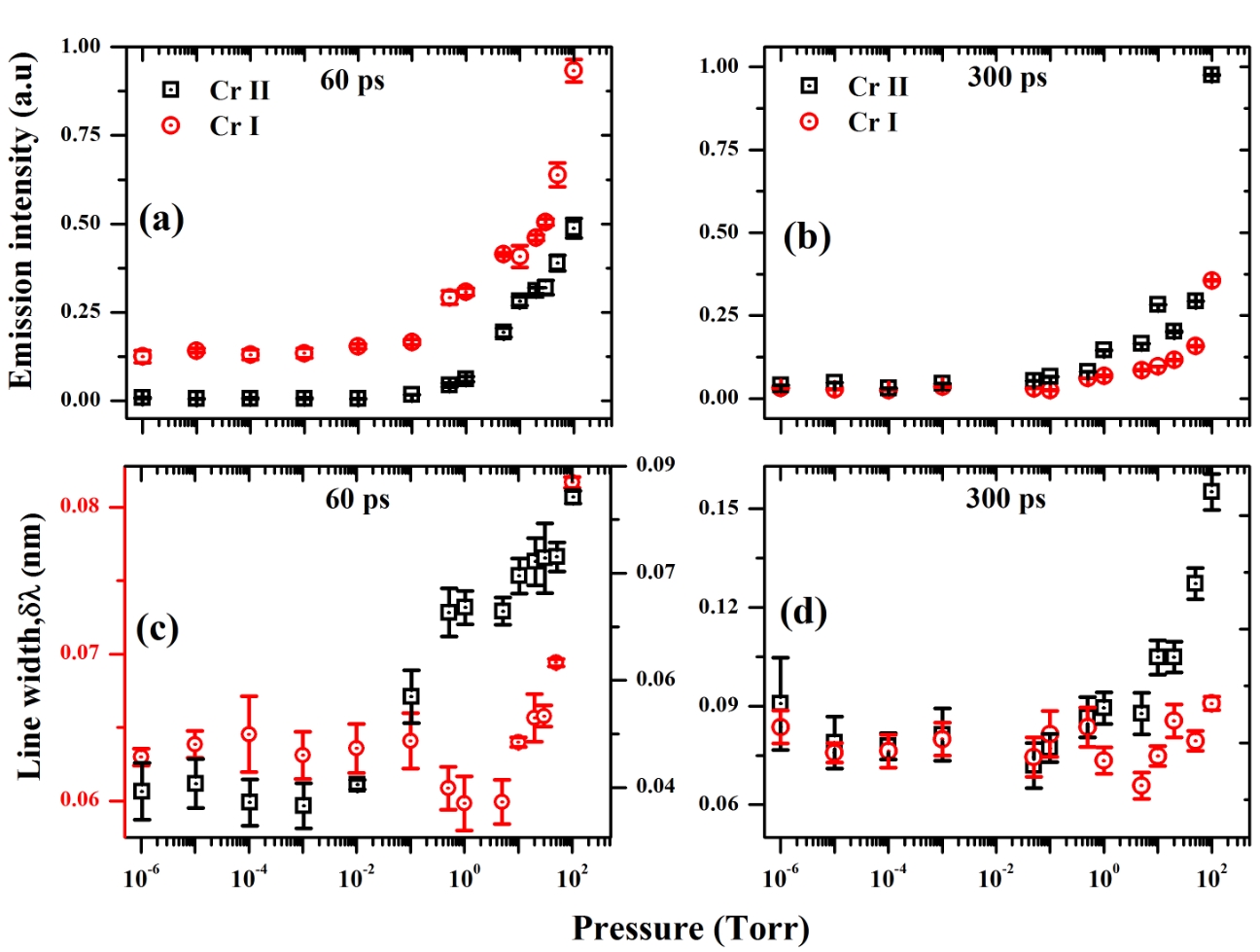}
\caption{Variation of normalized emission intensity and line width of Cr I and Cr II  with respect to ambient pressure for 60 ps and 300 ps laser irradiations. While the emission intensity and line width of ions are found to increase with an increase in ambient pressure for both irradiations (a and b), there is an oscillation for $\delta\lambda_{neutrals}$ for pressures between 0.5 Torr and 50 Torr (c and d). Measurements are carried out at $t_d$ = 50 ns and $t_w$ = 200 ns. Emission intensities are normalized using the maximum emission intensity in each case and error bars are obtained from multiple measurements.}\label{pressure Vs intensity and FWHM}
\end{figure}

To investigate the variation in $l_{EI}$, $\delta\lambda_{neutrals}$ and $\delta\lambda_{ions}$ in the  plume with respect to ambient pressures, OES measurements for both irradiations were performed and compared for $t_d$= 50 ns and $t_w$ = 200 ns. Results are given in figure \ref{pressure Vs intensity and FWHM}. $l_{EI}$ of neutrals and ions for both irratdiations remain unchanged in the range of 10$^{-6}$ Torr to 0.1 Torr, beyond which an increase is evident (figure \ref{pressure Vs intensity and FWHM} (a) and \ref{pressure Vs intensity and FWHM}(b)). At low pressures, an adiabatic expansion of the plume reduces collisional interactions among species causing a relatively lower excitation of the emitting entities, resulting in a lower $l_{EI}$\cite{smijesh2013OES, Harilal2003}. At intermediate pressures until $\sim$ 1 Torr, plume-background interaction and plasma confinement are the key factors deciding expansion. The plume front decelerates relatively quickly compared to that at lower pressures, causing enhanced collisional interactions among species and hence more charged species and excited species in the plume enhancing $l_{EI}$\cite{Andrey2000, Harilal2003, Harilal2014}. On a further increase of pressure, the interaction of plume with the background gets more complex due to various processes such as collisional excitations/de-excitations, formation of shock waves, plume splitting and sharpening \cite{Farid2014,Andrey2000,Harilal2003,Harilal2014} and thermal leak to the ambient\cite{smijesh2013OES, P.T.Rumsby1974}. 

Interestingly, $\delta\lambda_{neutrals}$ is found to have a distinctly different behaviour compared to $\delta\lambda_{ions}$ when the background pressure is increased from 10$^{-6}$ Torr to 10$^2$ Torr (figure \ref{pressure Vs intensity and FWHM} (c) and \ref{pressure Vs intensity and FWHM}(d)). $\delta\lambda_{ions}$ remains unchanged up to 0.1 Torr, but increases with an increase of pressure. This is in agreement with previous reports\cite{Farid2013, Farid2014} of an increase in $\delta\lambda$ with ambient pressure implying that the density of the plume varies/increases with pressure according to equation \ref{linewidth approximation}. Contrary to this behaviour of $\delta\lambda_{ions}$, $\delta\lambda_{neutrals}$ slightly decreases up to around 5 Torr before it increases with increasing pressure. The current observation of the oscillation of $\delta\lambda_{neutrals}$ for pressures between 0.5 Torr and 50 Torr (which is absent for $\delta\lambda_{ions}$) is similar to the experimental results reported by Rao \textit{et.al.}\cite{rao2016} for an integrated optical emission spectroscopy measurement of a fs laser produced zinc plasma. In that report\cite{rao2016}, an oscillation of line width has been explained on the basis of the variation of average temperature ($T_e$) and number density ($n_e$) with respect to pressure given by the equation\cite{Kunze} $\delta\lambda \propto \frac{n_e}{T_e^x}$; which is an outcome of impact broadening. Therefore, to understand this observation further, $T_e$ is calculated using the Boltzmann Plot method{\cite{Boltzmann1,harilal2005spectroscopic,Boltzmann2}} assuming local thermodynamic equilibrium (LTE).

\begin{figure}[ht!]
\includegraphics[scale=0.4]{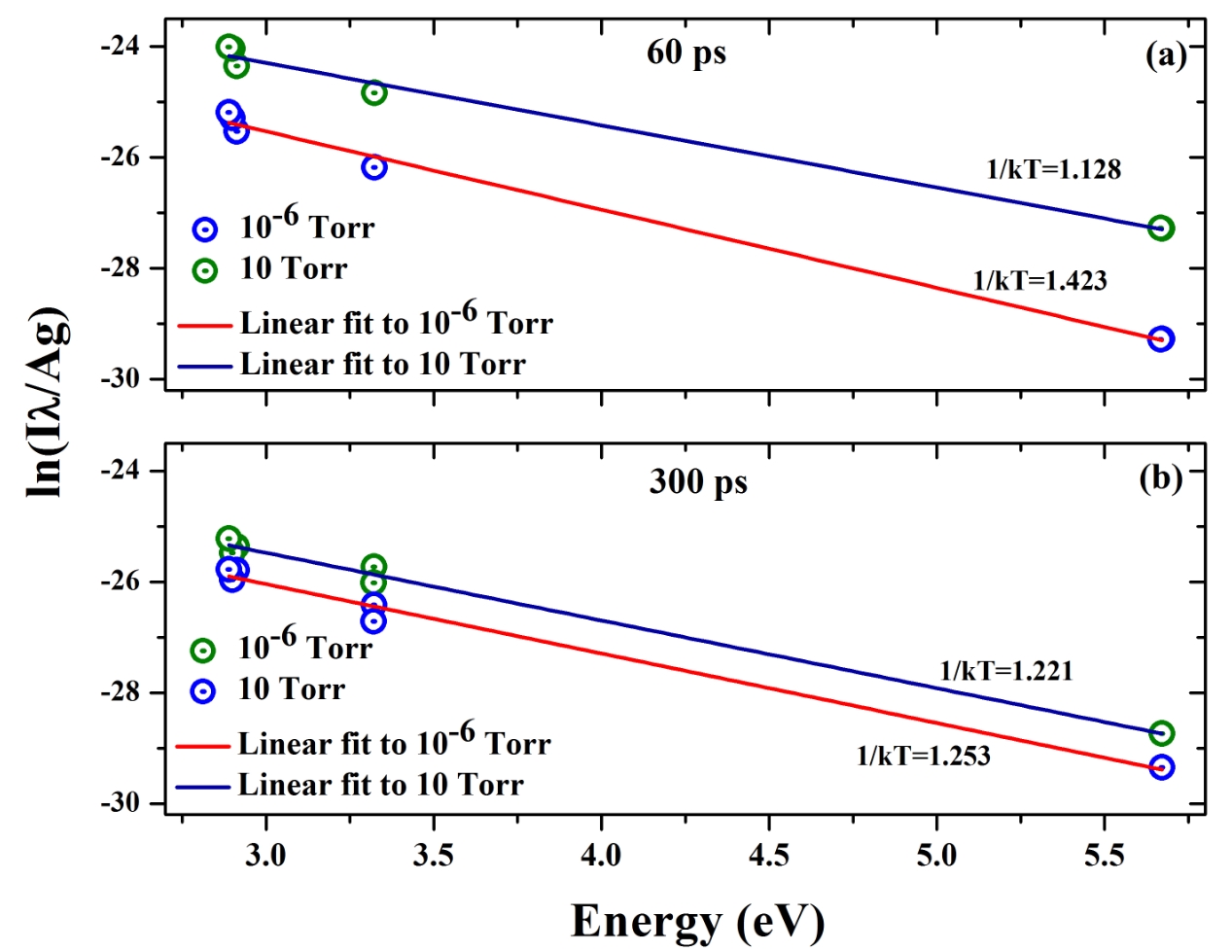}
\caption{Boltzmann plots (representative) used for the calculation of plasma temperature at 10$^{-6}$ Torr and 10 Torr for (a) 60 ps and (b) 300 ps irradiations. Emission lines from neutral chromium species (Cr I) are used for obtaining the Boltzmann plot.}\label{Boltzmann Plot}
\end{figure}

\begin{figure}[ht!]
\includegraphics[scale=0.4]{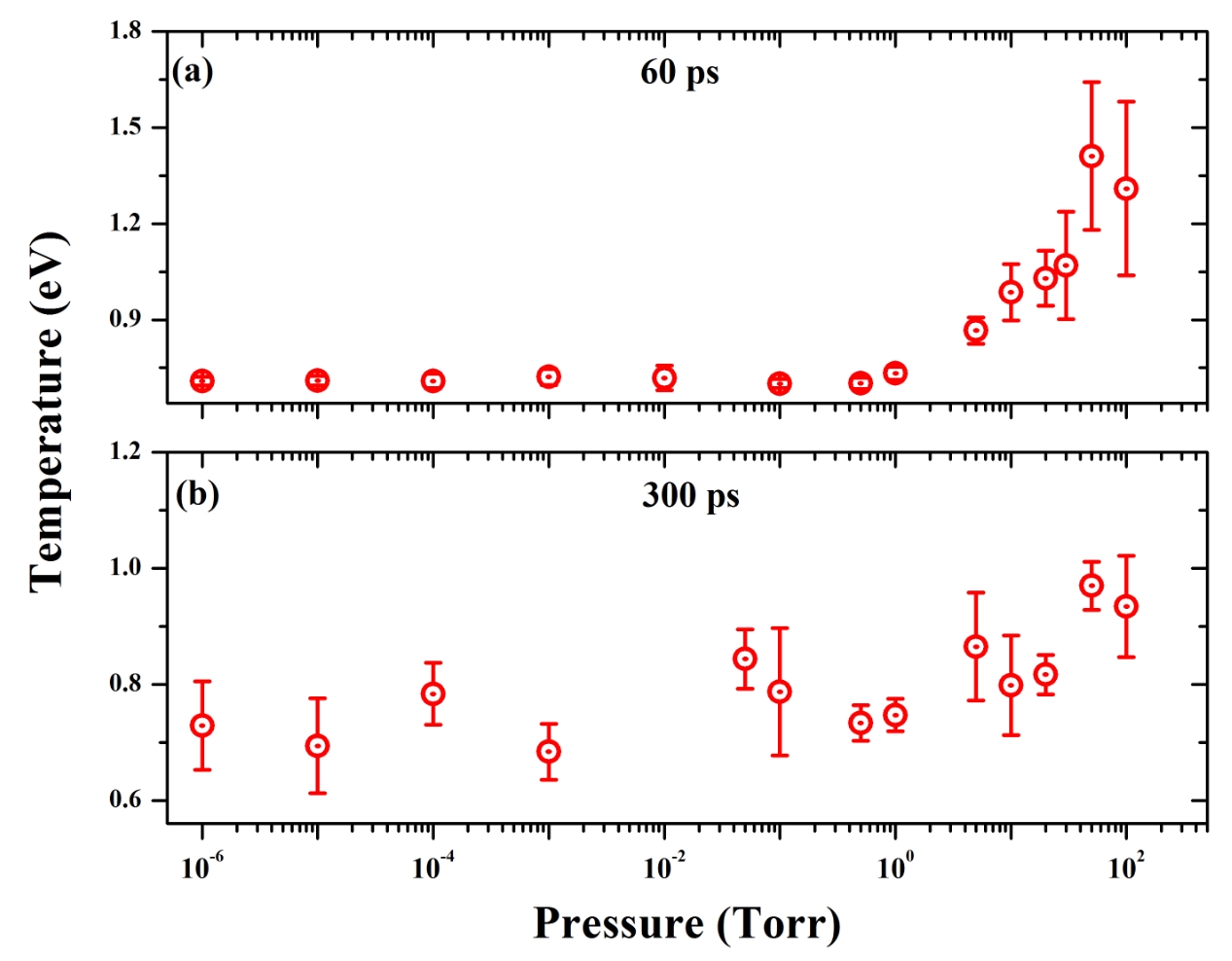}
\caption{Plasma temperature calculated for (a) 60 ps and (b) 300 ps irradiaitons using Boltzmann plot method assuming LTE for each pressure. Error bar is the standard deviation obtained throungh averaging multiple measurements.}\label{pressure vs temperature}
\end{figure}

For plasma in LTE, Boltzmann distribution gives the population of the excited state via\cite{griem2005principles,harilal2005spectroscopic}:
\begin{equation}
n_{nm}=n_n \frac{g_m}{Z} e^{-E_m/kT}
\end{equation}
where $n_{nm}$ is the population of the n$^{th}$ excited state, $n_n$ is the population of the lower level, $g_m$ is the statistical weight of the upper level of the transition, $E_m$ is the excitation energy, $k$ is the Boltzmann constant and $T$ is the temperature. Emission intensity is related to the population of the excited state by the equation\cite{griem2005principles}:
\begin{equation}
I_{nm}\approx A_{nm}n_{nm} \frac{hc}{\lambda_{nm}}=A_{nm}n_n \frac{g_m hc}{Z \lambda_{nm}}e^{-E_m/kT}
\end{equation}
where $A_{nm}$ is the atomic transition probability and $\lambda_{nm}$ is the wavelength of the line emission. 
Cr I line emissions at 396.36 nm, 396.97 nm, 425.45 nm, 427.49 nm, 428.97 nm, 520.60 nm and 520.84 nm were used to calculate the temperature (for every pressure), the result of which is given in figure \ref{pressure vs temperature}. $ln(I\lambda/Ag)$ versus energy ($E$) of the upper level of transitions is fitted to a straight line with a slope $-\frac{1}{kT}$ yields the temperature in equilibrium, the $T_e$. Boltzmann plots fitted for 60 ps and 300 ps irradiations at 10$^{-6}$ Torr and 10 Torr are given in figure \ref{Boltzmann Plot}. 
 
Figure \ref{pressure vs temperature} shows the variation of $T_e$ with respect to ambient pressure for 60 ps (figure \ref{pressure vs temperature} a) and 300 ps (figure \ref{pressure vs temperature} b) irradiations. $T_e$ has similar values when pressure is increased from 10$^{-6}$ Torr up to 0.5 Torr; but increases on further increase of pressure. The error bar  in $T_e$ beyond 5 Torr might have emanated from enhanced collisional interactions, within the plume due to plasma confinement upon increased ambient pressure levels. The process will be much more complex if the pulse is longer. For example, in the case of 300 ps irradiation, the temperature changes from $\sim$ 0.7 eV to $\sim$ 1 eV upon increasing pressure, and the large error bar makes it difficult to draw any conclusion though $T_e$ shows an increase with pressure. Such variations in temperature with large error bar can occur when the plasma plume slightly deviates from the LTE conditions. In addition, laser-plasma interaction changes the number density of the plume and hence reduces the number of neutrals which is important for 300 ps when compared to 60 ps, the reason for which has already been explained in the previous section. 

To understand the nature of oscillation of $\delta\lambda_{neutrals}$ for intermediate pressures for larger $T_e$, recall the equation \cite{Kunze,Baranger,Baranger2}:

\begin{equation}
\delta\lambda = n_e \int_0^\infty v f_e (v) \left[\sum\limits_{p'+p} \sigma_{pp'}(v) + \sum\limits_{q'+q} \sigma_{qq'}(v)\right] dv 
+ n_e \int_0^\infty v f_e (v) \left[\int |\phi_p (\theta,v) - \phi_q (\theta,v)|^{2} d\Omega\right]dv
\end{equation}

where the first term is the contribution to $\delta\lambda$ from inelastic collisions (whose cross sections are denoted by $\sigma$) connecting the upper ($p$) and lower ($q$) levels with other perturbing levels ($p'$ and $q'$). Integration over electron velocity distribution ($f_e (v)$) yields the rate coefficients. The second term denotes the contribution from elastic collisions wherein $\phi_p (\theta, v$ and $\phi_q (\theta, v)$ are the elastic scattering amplitudes for the target ion in the upper and lower states respectively. Integration is carried out over the scattering angle $\theta$ and $d\Omega$ is the element of the solid angle. Any changes in $n_e$ results in the modification of $\delta\lambda$ since elastic and inelastic collisions are dependent on the $n_e$. Changes in $n_e$ is highly probable in the current case either via laser-plasma energy coupling or via collisional excitation. More precisely, enhanced contribution to $n_e$, if any, has to come from collisional interaction upon increase in pressure as laser-plasma energy coupling is minimal for 60 ps. Whereas $n_e$ would be relatively larger for 300 ps due to better laser-plasma energy coupling apart from collisional interactions, which is rather evident in OES measurement. Additionally, collisional interaction is enhanced with ambient pressure due to the larger temperature, leading to more frequent electron-neutral collisions than electron-ion collisions, since the collision frequency varies as ${T_e}^{1/2}$ for the former and as ${T_e}^{-3/2}$ for the latter. Hence, it can be concluded that the line width of neutrals drops to a minimum value for an optimum pressure irrespective of the material from which the plasma is generated, provided there are instantaneous velocity changes for a large number of species interacting with the emitter due to the higher plasma temperatures. 

\section{Conclusion}
In short, plasmas generated by irradiating $\sim$ 0.60 mJ, $\sim$ 60 ps and 1 mJ, 300 ps laser pulses onto a solid Cr target is characterized using time-resolved optical emission spectroscopy. It is observed that the plasma generated by 60 ps irradiation resembles a fs LPP more than plasma generated by 300 ps irradiation, which is attributed to the interaction of the laser pulse with the plasma. Interestingly, an oscillation observed in the line width of neutrals ($\delta\lambda_{neutrals}$) for a pressure range from 0.5 Torr to 50 Torr, similar to a previous work\cite{rao2016} in femtosecond laser produced Zn plasma, is explained and correlated to collisional effects and changes in the instantaneous velocities of species interacting with the emitter. This is investigated experimentally by analysing the variation of temperature with respect to pressure for both irradiations. Enhanced collisions at larger plasma temperatures result in instantaneous velocity change of species interacting with emitters within the plume causing the line to broaden to its minimum value for neutrals since collisional frequency of neutrals in plasma is proportional to $T_e^{1/2}$. On the other hand collision frequency for ions is proportional to $T_e^{-3/2}$ and hence a monotonous increase for $\delta\lambda_{ions}$ is seen. Therefore the mechanisms causing oscillations for $\delta\lambda_{neutrals}$ would get modified due to the generation of excited species/ions via laser-plasma energy coupling and collisional interactions. The generation of more and more ions in the plume reduces the oscillatory nature of $\delta\lambda_{neutrals}$, which is evident in the current experiment.The observed oscillation of $\delta\lambda_{neutrals}$ is found to be irrespective of pulse duration (aleast for any pulse duration $\leq$ 300 ps) and target materials.
\begin{acknowledgments} 
This project is funded by the Australian Research Council linkage project grant No. LP140100813\cite{Linkagegrant}.
Kavya H. Rao has been supported through an ``Australian Government Research Training Program Scholarship".
\end{acknowledgments}
\bibliography{aipsamp}

%
%

%

\end{document}